\documentclass[11pt]{article}
\usepackage[margin=1in]{geometry}

\usepackage[utf8]{inputenc} 
\usepackage[T1]{fontenc}    
\usepackage{hyperref}       
\usepackage{url}            
\usepackage{booktabs}       
\usepackage{amsfonts}       
\usepackage{nicefrac}       
\usepackage{microtype}      
\usepackage{xcolor}         
\usepackage{combelow}
\usepackage{amsmath, amssymb, amsfonts}
\usepackage{bm}  
\usepackage{tikz}

\usepackage{algorithm}
\usepackage{algpseudocode}
\usepackage{array}
\usepackage{caption}
\usepackage{wrapfig}

\usepackage{subcaption}
\usepackage{graphicx}
\usepackage{makecell}
\usepackage[numbers]{natbib}

\newcommand{\eg}{\emph{e.g.},\ }
\newcommand{\ie}{\emph{i.e.},\ }

\DeclareMathOperator*{\argmin}{arg\,min}

\title{
\Large \bfseries
\textbf{GIFT}: Gradient-aware Immunization \\of diffusion models against malicious Fine-Tuning \\with safe concepts retention 
\rule{\textwidth}{0.7pt}
}

\author{
  Amro Abdalla\textsuperscript{1} \quad
  Ismail Shaheen\textsuperscript{1} \quad
  Dan DeGenaro\textsuperscript{1} \quad
  Rupayan Mallick\textsuperscript{1} \\
  Bogdan Rai\cb{t}\u{a}\textsuperscript{2} \quad
  Sarah Adel Bargal\textsuperscript{1} \\
  \textsuperscript{1}Department of Computer Science, Georgetown University \\
  \textsuperscript{2}Department of Mathematics and Statistics, Georgetown University \\
  \texttt{\{aaa654, ias68, drd92, rupayan.mallick, br607, sarah.bargal\}@georgetown.edu}
}
\date{} 

\begin{document}

\maketitle


\begin{abstract}
We present $\textbf{GIFT}$: a $\textbf{G}$radient-aware $\textbf{I}$mmunization technique to defend diffusion models against malicious $\textbf{F}$ine-$\textbf{T}$uning while preserving their ability to generate safe content. Existing safety mechanisms like safety checkers are easily bypassed, and concept erasure methods fail under adversarial fine-tuning. GIFT addresses this by framing immunization as a bi-level optimization problem: the upper-level objective degrades the model’s ability to represent harmful concepts using representation noising and maximization, while the lower-level objective preserves performance on safe data. GIFT achieves robust resistance to malicious fine-tuning while maintaining safe generative quality. Experimental results show that our method significantly impairs the model’s ability to re-learn harmful concepts while maintaining performance on safe content, offering a promising direction for creating inherently safer generative models resistant to adversarial fine-tuning attacks. \\
{\small\textcolor{red}{Warning: This paper contains NSFW content.  Reader discretion is advised.}}
\end{abstract}

\section{Introduction}




Text-to-image (T2I) models have emerged as powerful generative tools capable of producing high-quality images faithful to input prompts \cite{Rombach_2022_CVPR, podell2023sdxlimprovinglatentdiffusion, imagenteamgoogle2024imagen3, dalle_ramesh_2021}. However, their accessibility and adaptability make them vulnerable to malicious fine-tuning, where adversaries adapt pre-trained models to generate harmful or copyrighted content. Methods like DreamBooth \cite{Ruiz_2023_CVPR}, LoRA \cite{hu2022lora} , and Textual Inversion \cite{gal2023an} enable this adaptation with minimal resources and without needing to train from scratch. This vulnerability persists even when existing safety mechanisms, such as safety checkers \cite{rando2022red} or concept erasure methods \cite{gandikota2023erasing, gandikota2024unified, rece_chao_gong}, are in place, as they can be bypassed \cite{yang_sneaky_2024, zhang2024generate, gao2024htsattackheuristictokensearch}, disabled, or undone through lightweight adaptation techniques. This creates a significant risk: once a model is open-sourced, it becomes difficult to guarantee its continued alignment with safety goals. Current defenses either degrade the model’s generative capabilities or fail to withstand adversarial fine-tuning.



While safety checkers and licensing agreements offer a first line of defense \cite{compvis2022stable}, they are not an inherent property of T2I models and are easily circumvented \cite{rando2022red}. To enhance the inherent safety of T2I models, concept erasure techniques have been proposed to remove undesirable concepts by modifying the model's internal representations. Although such techniques can suppress the generation of undesired concepts, they are vulnerable to circumvention~\cite{pham2024circumventing, zhang2024generate}. Moreover, as we show in our experiments, simple fine-tuning can reintroduce the erased concepts, undermining the long-term effectiveness of concept erasure methods as a safety mechanism.


To address the limitations of concept erasure and defend against its circumvention, model immunization has been proposed as a proactive defense against malicious fine-tuning of T2I models. IMMA~\cite{zheng2024imma}, for example, introduces a bi-level optimization approach inspired by MAML~\cite{finn2017model}, aiming to learn poor model initializations that hinder adaptation to undesirable concepts. By simulating the fine-tuning process during immunization, IMMA makes it more difficult for adversaries to reintroduce harmful content through fine-tuning. However, IMMA’s framework significantly compromises the model’s performance on safe concepts, degrading both its generative quality and its ability to be fine-tuned for benign applications as we will show in our experiments.

To this end, we propose \textbf{GIFT}—a \textbf{G}radient-aware \textbf{I}mmunization framework to defend T2I diffusion models against malicious \textbf{F}ine-\textbf{T}uning while preserving their ability to generate safe content. Inspired by IMMA~\cite{zheng2024imma} and MAML~\cite{finn2017model}, we formulate GIFT as a bi-level optimization problem: the lower-level task minimizes a \textit{prior preservation} objective to retain performance on safe concepts, while the upper-level task minimizes an \textit{immunization} objective that prevents adaptation to harmful concepts. The \textit{immunization} objective is composed of two parts: (1) a \textit{loss maximization term} and (2) a \textit{representation noising term} inspired by~\cite{rosati2024representation} from the language domain.


We demonstrate that immunizing a model with GIFT significantly impairs its ability to re-learn harmful content while maintaining generative ability across a wide range of safe concepts. Our evaluation covers several concept categories—including objects, art styles, and NSFW content—and considers multiple fine-tuning strategies, \eg LoRA and DreamBooth. Our main contributions are:
\begin{itemize}
    \item We propose GIFT, a novel framework that immunizes text-to-image diffusion models against malicious fine-tuning while preserving their generative quality and utility on safe concepts.
    \item We formulate immunization as a bi-level optimization problem where the lower-level task uses a prior preservation loss to maintain generation quality on safe concepts, and the upper-level task employs an immunization loss to resist adaptation to harmful ones.
    \item We conduct extensive experiments across diverse concept types (objects, art styles, and NSFW content), demonstrating that GIFT outperforms existing baselines (ESD, IMMA) in resisting malicious fine-tuning while preserving safe model utility.
\end{itemize}

\section{Related Work}
\label{sec:related_works}
The advancements of text-to-image (T2I) generative models, such as Stable Diffusion~\cite{Rombach_2022_CVPR}, have democratized content creation but also introduced significant risks associated with their misuse. A key concern is the vulnerability of these models to malicious fine-tuning, where bad actors can adapt pre-trained models to generate harmful, copyrighted, or otherwise undesirable content, prompting growing interest in developing safeguards for T2I models.

Existing approaches to mitigate these risks can be broadly categorized. One line of work focuses on \textbf{concept erasure} or unlearning, aiming to remove specific concepts from a pre-trained model. Erased Stable Diffusion (ESD)~\cite{gandikota2023erasing} fine-tunes model weights using textual descriptions of the undesired concept to prevent the model from generating it. Other methods explore unlearning by modifying specific model components like the text encoder or attention layers~\cite{kumari2023conceptablation, Zhang_2024_CVPR}, sometimes using few-shot unlearning techniques~\cite{wu2025unlearning}, by adding lightweight eraser modules~\cite{rece_chao_gong}, or through data unlearning ~\cite{alberti2025data}.

While effective at removal, some erasure techniques can be circumvented by further fine-tuning~\cite{pham2024circumventing}, as the underlying knowledge might not be entirely eliminated or can be easily relearned~\cite{Zhou2024OnTL, zhang2024generate}. Additionally, a significant challenge is preserving the model's utility on unrelated concepts, as aggressive erasure can lead to ``catastrophic forgetting'' of desired knowledge~\cite{tian-etal-2024-forget, Xu_MU_Survey}. Some recent works attempt to address this by focusing on concept-localized regularization or mitigating conflicting gradients during unlearning~\cite{patel2025learningunlearnretainingcombating, wu2025unlearning}.

Another paradigm is \textbf{model immunization}, which seeks to make the model inherently resistant to adaptation towards malicious concepts before it is released. IMMA (Immunizing text-to-image Models against Malicious Adaptation)~\cite{zheng2024imma} proposes learning model parameters that are difficult for adaptation methods to fine-tune on malicious content, framed as a bi-level optimization problem. While IMMA demonstrates effectiveness against various adaptation methods like LoRA~\cite{hu2022lora}, Textual Inversion~\cite{gal2023an}, and DreamBooth~\cite{Ruiz_2023_CVPR}, it can be overly aggressive, potentially degrading the model's performance on safe, unrelated concepts. Other defense strategies include methods akin to data poisoning (\eg Glaze~\cite{shawn_2023}, which protects artistic styles from mimicry).

Other defense approaches focus on \textbf{safe decoding or generation}, often by modifying the diffusion process~\cite{schramowski2023safe} or employing external classifiers and adaptive guards to filter outputs~\cite{yoon2025safree, wang2024aeiouunifieddefenseframework}. However, these can sometimes be bypassed by users with white-box access to the model or through carefully crafted adversarial prompts and jailbreaking methods~\cite{rando2022red, yang_sneaky_2024, gao2024htsattackheuristictokensearch, liu2024jailbreak}.

Techniques from the Large Language Model (LLM) domain are also being explored and adapted. Representation Noising (RepNoise)~\cite{rosati2024representation}, for instance, has been proposed as a defense mechanism against harmful fine-tuning in LLMs by removing information about harmful representations across model layers, making them difficult to recover. GIFT draws inspiration from this by adapting representation noising to T2I models.

Unlike some erasure methods that can be easily reversed or circumvented~\cite{pham2024circumventing, zhang2024generate}, and in contrast to immunization methods like IMMA~\cite{zheng2024imma} that may overly degrade general utility, GIFT aims for a better trade-off. GIFT's bi-level formulation helps coordinate the optimization to prevent the immunization objective from detrimentally affecting the prior preservation objective.
\section{Methodology}
\label{sec:methodolgy}
\subsection{Problem Formulation}

Our goal is to prevent adaptation methods from reintroducing malicious concepts into a pre-trained T2I diffusion model. We formulate this as a bi-level optimization problem with two objectives: (1) immunization against harmful concepts and (2) preservation of model performance on safe data.

Let \(\theta\) represent the U-Net parameters of a pre-trained T2I model (\eg Stable Diffusion \cite{Rombach_2022_CVPR}), and \(\psi \subset \theta\) denote the subset of parameters corresponding to cross-attention layers. Let \((x_m, c_m) \in D_M\) and \((x_s, c_s) \in D_S\) denote image–concept pairs from the malicious and safe datasets, respectively. We aim to derive an immunized model \(\theta^I\) that resists adaptation to malicious concepts under any subsequent fine-tuning while maintaining its ability to learn and generate safe concepts. 

\subsection{Bi-level Optimization Framework}\label{sec:bi-level}

The authors of IMMA \cite{zheng2024imma} employ a meta-learning algorithm inspired by MAML \cite{finn2017model} to immunize T2I models. They simulate malicious adaptation steps by minimizing the adaptation loss in the lower-level task, while maximizing the same loss in the upper-level task to achieve immunization. We employ a similar framework to immunize a T2I model against malicious concepts while retaining performance on safe data. We define the upper-level task as the immunization objective over \(D_M\) and the lower-level task as the prior preservation objective over \(D_S\). We show our algorithm at \ref{alg:immunization}.

To perform optimization, we compute parameters $\theta^*$ via a gradient step using the lower-level task on \(D_S\), followed by an optimization step of said parameters using the upper-level task on \(D_M\). We formulate this as the following bi-level optimization problem:
\begin{align}\label{eq:bi-level}
\underbrace{\psi^I = \argmin_{\psi \subset \theta^*} \mathcal{L}_\text{immunize}(x_m, c_m; \theta^*)}_{\text{upper-level task}} \quad \text{where} \quad \underbrace{\theta^* = \argmin_{\theta} \mathcal{L}_\text{prior}(x_s, c_s; \theta)}_{\text{lower-level task}} .
\end{align}
In the upper-level task, we minimize the \textit{immunization loss} \(\mathcal{L}_\text{immunize}\) with respect to $\psi$ (\ie cross-attention layers). This encourages the model to resist adapting to harmful concepts from the malicious dataset \(D_M\).
In the lower-level task, we minimize the \textit{prior preservation loss} $\mathcal{L}_\text{prior}$ with respect to the U-Net parameters $\theta$, which includes $\psi$. This selection reflects the intuition that cross-attention layers play a central role in encoding and manipulating concepts \cite{Liu_2024_CVPR}, while optimizing the U-Net in the inner loop ensures that the model adapts safely while incorporating the immunization updates.


To further explore how this bi-level setup benefits the immunization approach more than naive addition of all losses, we examine intermediate gradient updates. We implement the bi-level scheme in Eq.~\eqref{eq:bi-level} by iterating the gradient updates:
\begin{equation}
    \theta'=\theta-\alpha_P\nabla\mathcal L_P(\theta)\quad\text{and}\quad\psi''=\psi'-\alpha_I\nabla \mathcal L_I(\psi'),
\end{equation}
where \(\psi'\subset\theta'\), and we abbreviate \(\mathcal L_\mathrm{prior}(x_s,c_s;\theta)=\mathcal L_P(\theta)\) and \(\mathcal L_\mathrm{immunize}(x_m,c_m;\theta)=\mathcal L_I(\psi)\). We then use the Taylor Series expansion for \(\nabla \mathcal L_I(\psi')\) as seen in Eq~\eqref{eq:taylor}, which gives us the total update for $\psi''$ in Eq~\eqref{eq:grad-update}. Full derivation can be found in the appendix.
\begin{equation}\label{eq:taylor}
\nabla \mathcal L_I(\psi')\approx\nabla\mathcal L_I(\psi)-\alpha_P\nabla^2 \mathcal L_I(\psi)\nabla_\psi \mathcal L_P(\theta)
\end{equation}
\begin{equation} \label{eq:grad-update}
    \psi''\approx \psi -\textcolor{blue}{\alpha_P\nabla_\psi \mathcal L_P(\theta)}-\textcolor{orange}{\alpha_I\nabla \mathcal L_I(\psi)}+\textcolor{purple}{\alpha_P\alpha_I\nabla^2 \mathcal L_I(\psi)\nabla_\psi \mathcal L_P(\theta)}
\end{equation}
The final equation shows that our current immunization gradient update is equivalent to doing an update in the \textcolor{blue}{\textit{prior preservation}} direction plus an update in the \textcolor{orange}{\textit{immunization}} direction plus an \textcolor{purple}{\textit{additional}} term. That term is the directional curvature of $\mathcal L_I$ along $\nabla_\psi \mathcal L_P(\theta)$, which is crucial for our approach. This term adds a second order correction which helps coordinate the gradient descent so that minimizing \(\mathcal L_I\) does not make minimizing $\mathcal L_P$ harder. This improves the retention of our model significantly by making the immunization update ``aware'' of previous prior preservation updates.

\subsection{Immunization Loss}

The immunization loss that we employ in the upper-level task consists of two components: (1) loss maximization and (2) representation noising.

\textbf{Loss Maximization.}
We maximize the loss with respect to the malicious concept as follows: 
\begin{equation}
\mathcal{L}_\text{max}  = -\mathbb{E}_{t, \epsilon \sim \mathcal{N}(0,I)} \left[ \|\epsilon_\theta(x_m, c_m, t) - \epsilon\|_2^2 \right] .
\label{eq:max}
\end{equation}
This maximization aims to push the model parameters $\theta$ to perform poorly on the target malicious data \((x_m, c_m)\). However, loss maximization on malicious content is not sufficient on its own.

\textbf{Representation Noising.}
While maximizing the loss on malicious concepts reduces the model’s ability to generate them, it does not necessarily prevent the model from re-adapting to these concepts with further fine-tuning. This is because when maximizing $\mathcal{L}_\text{max}$, the mutual information $\text{MI}(x_m|c_m;y_m)$ between conditioned malicious inputs \(x_m|c_m\) and malicious model outputs \(y_m\) is targeted, but the mutual information $\text{MI}(x_m|c_m;z_m)$ between inputs \(x_m|c_m\) and intermediate representations \(z_m\) can remain, which may allow the malicious concept to return \cite{rosati2024representation}. The data processing inequality states:
\begin{equation}
    \text{MI}(x_m|c_m;z_m) \geq \text{MI}(x_m|c_m;y_m) .
\end{equation}
That is, information shared between inputs $x_m|c_m$ and intermediate representations $z_m$ is an upper bound on information shared between those inputs and the outputs $y_m$. As such, it is useful to directly reduce $\text{MI}(x_m|c_m;z_m)$ which implies a reduction in $\text{MI}(x_m|c_m;y_m)$. To this end, we adapt the LLM immunization technique of \cite{rosati2024representation}, representation noising, to T2I models. Let \( L^{(j)} \) denote the \( j \)-th layer of the U-Net, where \( j \in \{1, \ldots, n\} \). 

For an input to the U-Net conditioned on the malicious concept $x_m|c_m$, we define the first intermediate representation as $z_m^{(1)} = L^{(1)}(x_m|c_m)$ and further intermediate representations as \( z_m^{(j)} = L^{(j)}(z_m^{(j-1)}) \) for $j \in \{2, \ldots, n\}$. We then minimize the loss between these activations and random noise:
%
\begin{equation}
\mathcal{L}_\text{noise} = \textstyle{\sum_{j=1}^n} \text{MSE} \left( z_m^{(j)}, \epsilon_m^{(j)} \right), \quad \text{where } \epsilon_m^{(j)} \sim \mathcal{N}\left(\mu_{z_m^{(j)}}, \sigma^2_{z_m^{(j)}}\right) ,
\label{eq:noise}
\end{equation}
where \(\left(\mu_{z_m^{(j)}}, \sigma^2_{z_m^{(j)}} \right)\) are the computed mean and variance of the sample $z_m^{(j)}$. This loss aims to destroy undesirable information contained in the model while preventing model parameters from diverging significantly from their pre-training.


\textbf{Total Immunization Loss.} The final immunization objective combines the loss maximization term with the representation noising loss, weighted by a hyperparameter $\beta$:
\begin{equation}
\mathcal{L}_\text{immunize} = \mathcal{L}_\text{max} + \beta \cdot \mathcal{L}_\text{noise} .
\end{equation}
This loss is applied specifically to the cross-attention layers in the upper-level optimization to target the parts of the model most responsible for concept encoding.

\subsection{Prior Preservation Loss}
The immunization loss can degrade the model's performance on safe tasks. To mitigate this effect, we employ the original T2I model training objective for safe data preservation:
\begin{equation}
\mathcal{L}_\text{prior}  = \mathbb{E}_{t, \epsilon \sim \mathcal{N}(0,I)} \left[ \|\epsilon_\theta(x_s, c_s, t) - \epsilon\|_2^2 \right],
\label{eq:prior}
\end{equation}
which helps to maintain performance on safe concepts while immunizing against malicious ones. 
\begin{algorithm}[t]
\caption{Our method (GIFT)}
\label{alg:immunization}
\begin{algorithmic}[1]
\Require Malicious dataset $D_M$, safe dataset $D_S$ 
\Require Model parameters $\theta$ with cross-attention subset $\psi \subset \theta$
\Require Learning rates $\alpha_\text{inner}$, $\alpha_\text{outer}$, Noising weight $\beta$ 
\For{each training iteration}
    \If{inner loop step}
        \State Sample batch $(x_s, c_s)$ from $D_S$ \Comment{Lower-level task: Prior Preservation}
        \State \(\theta \leftarrow \theta - \alpha_\text{inner} \nabla_\theta \mathcal{L}_\text{prior}(x_s, c_s; \theta)\)
    \Else
        \State Sample batch $(x_m, c_m)$ from $D_M$ \Comment{Upper-level task: Immunization}
        \State $\mathcal{L}_\text{max} \leftarrow -\mathbb{E}_{t, \epsilon \sim \mathcal{N}(0,I)} \left[ \|\epsilon_\theta(x_m, c_m, t) - \epsilon\|_2^2 \right]$
        \State Extract intermediate activations $z_m^{(j)}$ for layers $j=1,\dots,n$
        \State Sample noise $\epsilon_m^{(j)} \sim \mathcal{N}\left(\mu_{z_m^{(j)}}, \sigma^2_{z_m^{(j)}} \right)$ \Comment{Mean and Var. from $z_m^{(j)}$}
        \State $\mathcal{L}_\text{noise} \leftarrow \sum_{j=1}^n \text{MSE} \left( z_m^{(j)}, \epsilon_m^{(j)} \right)$
        \State $\mathcal{L}_\text{immunize} \leftarrow \mathcal{L}_\text{max} + \beta \cdot \mathcal{L}_\text{noise}$
        \State $\psi \leftarrow \psi - \alpha_\text{outer} \nabla_\psi \mathcal{L}_\text{immunize}$
    \EndIf
\EndFor
\State \Return Immunized model parameters $\theta$ as $\theta^I$ 
\end{algorithmic}
\end{algorithm}









\section{Experiments}
\label{sec:experiments}
In this section, we show GIFT's ability to immunize the T2I model Stable Diffusion v1.5 (SD) \cite{Rombach_2022_CVPR} on objects, art styles, and NSFW content. We evaluate our method against IMMA and ESD.

\textbf{Experimental Setup.} For object immunization, we select 26 objects from the Custom Concept 101 dataset~\cite{Kumari_2023_CVPR}, each with more than 8 images split into 2 disjoint sets: \(D_M\) and \(D_A\). The defense (malicious) split \(D_M\) is used during immunization, and the attack split \(D_A\) is used to simulate malicious fine-tuning with DreamBooth. For prior preservation, we generate 500 safe images per object using category-level prompts to form a safe set, \(D_S\). For each object, we compare GIFT to IMMA and an undefended baseline. Similarly, for artistic styles, we test on 10 styles (\eg Van Gogh, Picasso) by generating 40 images per artist with prompts like \texttt{<a painting in [artist] style>}, splitting them equally into disjoint \(D_M\) and \(D_A\) sets. We compare GIFT to ESD and IMMA and we use \texttt{<a painting of a cat in [artist] style>} as a validation prompt. Finally, for NSFW content, we use the \texttt{porn} subset of the NSFW-T2I dataset~\cite{zxbsmk_nsfw_t2i}, sampling 40 images and dividing them into \(D_M\) and \(D_A\) sets. We used a single NVIDIA L40 GPU with 40GB of memory in an internal cluster for each experiment.

\textbf{Evaluation Metrics.}
We evaluate GIFT using four metrics: CLIP similarity for prompt-image alignment \cite{Hessel2021CLIPScoreAR}, LPIPS for perceptual fidelity \cite{zhang2018unreasonable}, DINO similarity for feature-level consistency \cite{Ruiz_2023_CVPR}, and NudeNet \cite{bedapudi2019nudenet} to quantify explicit content after immunization. Together, these capture semantic alignment, visual quality, safe concept retention, and NSFW suppression.

\subsection{Objects}
\label{sec:objects}
\begin{figure}[t]
    \centering
    \includegraphics[width=1\linewidth]{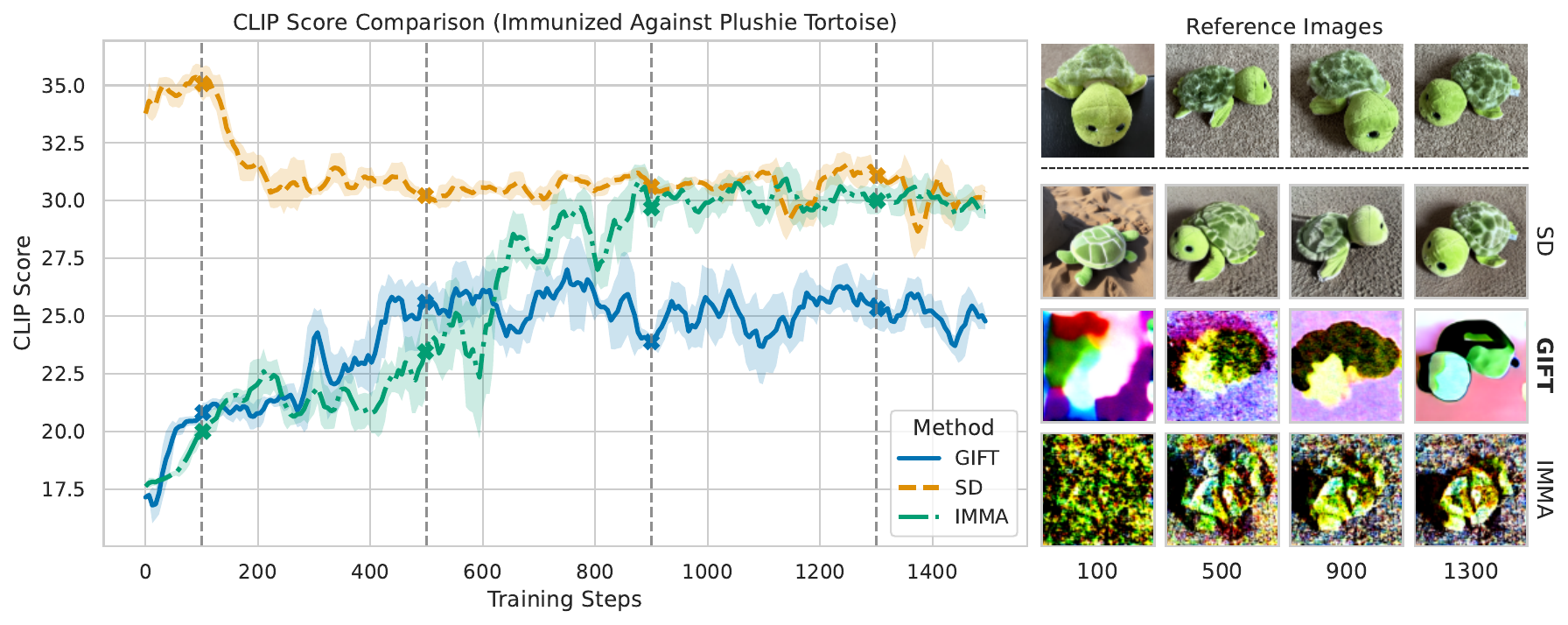}
    \caption{{\bf GIFT Immunizes Similarly to IMMA.} Here, we treat the tortoise plushie as a malicious concept using the prompt \texttt{<a *s tortoise plushie on the beach>} where \texttt{*s} is DreamBooth's special token. \textbf{Top row:} Reference images used to fine-tune via DreamBooth. \textbf{Second row:} Results of fine-tuning the undefended SD. \textbf{Third row:} Results of fine-tuning after 1K steps of immunization with GIFT. \textbf{Bottom row:} Results of fine-tuning after 1K steps of immunization with IMMA.}
    \label{fig:tortoise_atk}
\end{figure}

\begin{figure}[t]
    \centering
    \includegraphics[width=1\linewidth]{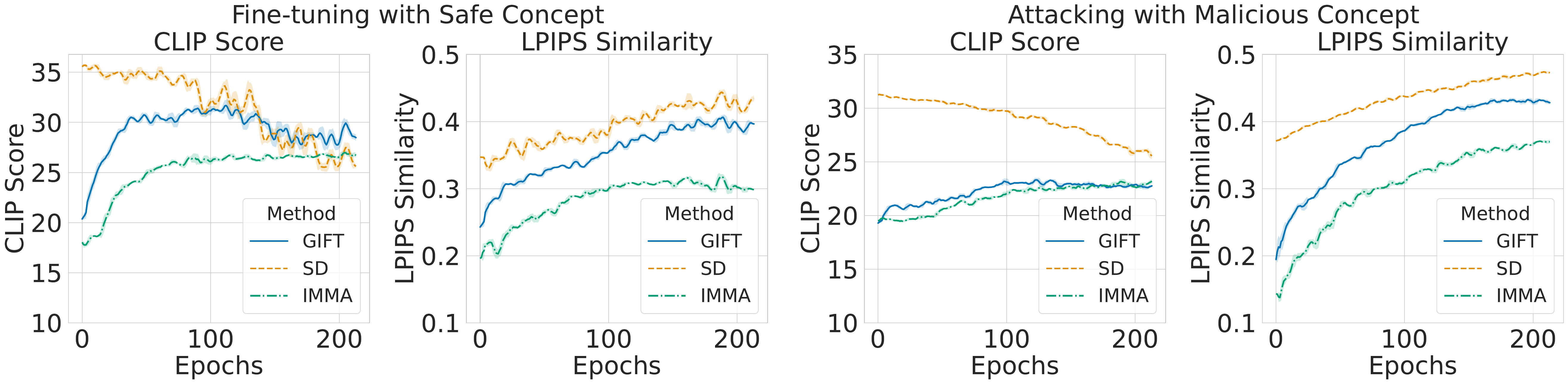}
    \caption{{\bf GIFT Finds a Middle Ground.} Averaged per-epoch CLIP Score and LPIPS Similarity across 26 immunized models during fine-tuning. GIFT achieves significantly higher CLIP and LPIPS similarities between images of the safe concept and its corresponding prompt than models immunized with IMMA, indicating preservation of generative capabilities. It additionally achieves similar CLIP scores to IMMA on malicious concepts, and significantly lower LPIPS scores than the undefended model, indicating successful immunization. Qualitatively, as seen in Figure~\ref{fig:tortoise_atk}, GIFT's scores still indicate sufficient immunization.}
    \label{fig:obj_avg}
\end{figure}

\textbf{Attack Results.}
We find that GIFT performs similarly to IMMA in terms of immunizing SD against particular concepts, and generally achieves CLIP and LPIPS scores ranging between those of the undefended model and those of a model defended with IMMA. In cases such as Fig.~\ref{fig:tortoise_atk}, GIFT outperforms IMMA by producing images with lower CLIP scores when prompted for the concept against which the model is immunized. Averaged per-epoch metrics across all 26 objects can be seen in Fig.~\ref{fig:obj_avg}. We do not view this overall quantitative difference in our results as compared with IMMA's as a weakness; rather, it indicates a less aggressive, but still functional immunization technique that preserves the model's generative capabilities to a great extent.

\begin{figure}[t]
    \centering
    \includegraphics[width=1\linewidth]{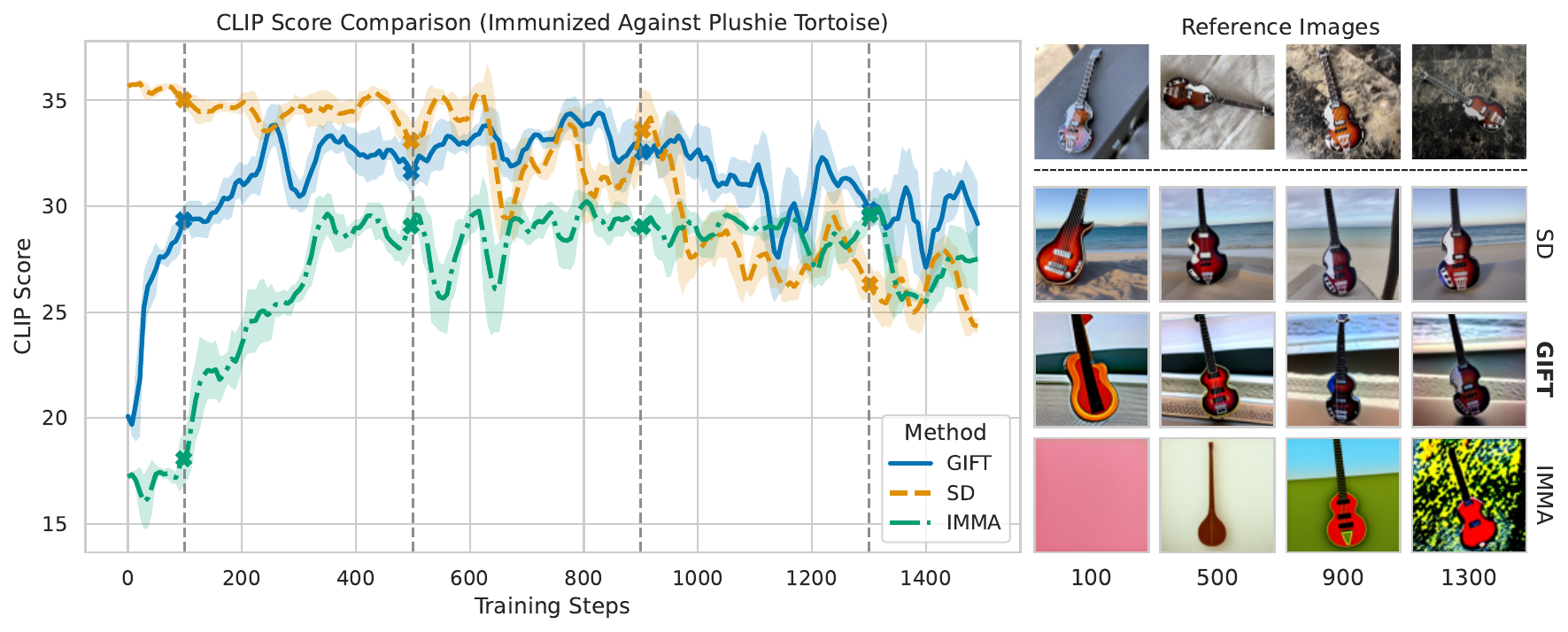}
    \caption{{\bf GIFT Retains Safe Concepts Better than IMMA.} Here, we treat the bass guitar as a safe concept unrelated to the malicious concept from Figure~\ref{fig:tortoise_atk} using the prompt \texttt{<a *s bass guitar on the beach>}. \textbf{Top row:} Reference images used to fine-tune via DreamBooth. \textbf{Second row:} Results of fine-tuning the undefended SD. \textbf{Third row:} Results of fine-tuning after 1K steps of immunization against the plushie from Figure~\ref{fig:tortoise_atk} with GIFT. \textbf{Bottom row:} Results of fine-tuning after 1K steps of immunization against the plushie from Figure~\ref{fig:tortoise_atk} with IMMA.}
    \label{fig:tortoise_ft}
\end{figure}

\textbf{Preservation Results.}
Models immunized with GIFT generally outperform those immunized with IMMA when tasked with generating images of a safe concept as can be seen qualitatively in Fig.~\ref{fig:tortoise_ft}. Models immunized with GIFT achieve CLIP and LPIPS scores similar to the undefended SD checkpoint. Averaged per-epoch metrics across all 26 objects can be seen in Fig.~\ref{fig:obj_avg}. Models immunized with IMMA generally achieve much lower similarity scores.

\subsection{Art Styles}
\label{sec:art-style}

\begin{figure}[t]
    \centering
    \includegraphics[width=1\linewidth]{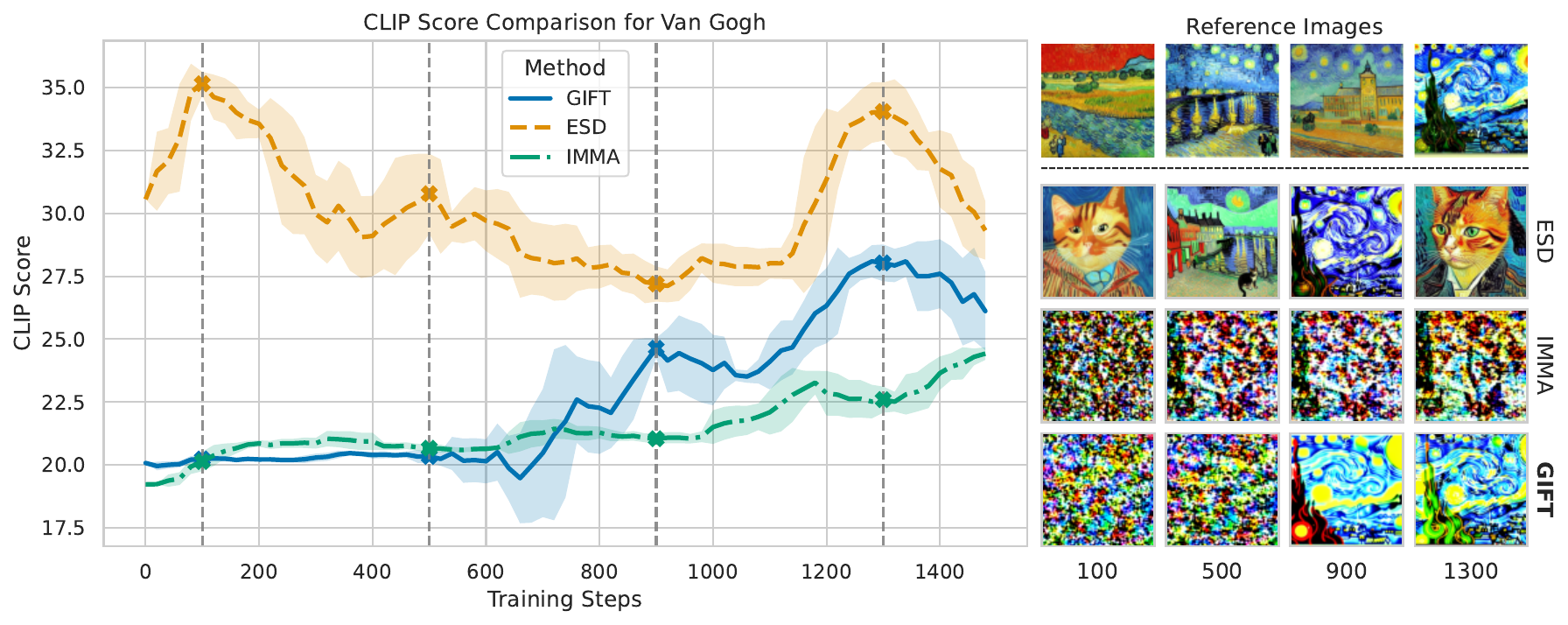}
    \caption{{\bf GIFT Prevents Artistic Style Adaption.} We fine-tune each model on a dataset of 20 Van Gogh generations (reference images included) and validate using the prompt \texttt{<a painting of a cat in [artist] style>}. On the left is the CLIP score for each method over the duration of training. On the right are qualitative results for each method at the 100, 500, 900, and 1300 step mark. ESD isn't able to prevent adaption to the protected art style. IMMA consistently produces noise for the protected model at the expense of degraded model performance. GIFT prevents the adaption to the protected art style by producing noise for the first part of the attack then overfitting at the end.}
    \label{fig:van-gogh-comparison}
\end{figure}

\paragraph{Attack Results.}
As shown in Fig.~\ref{fig:van-gogh-comparison}, ESD rapidly reacquires Van Gogh’s style (by step 100), including its application to unseen concepts, \eg cats. Then, the model enters a corruption phase, where overfitting becomes apparent. This is evidenced by a decline in CLIP score alongside increasing similarity to the training data. Near step 1300, we observe a transient improvement phase, followed by further degradation behavior consistent with previously observed fine-tuning dynamics in diffusion models~\cite{wu2024exploringdiffusionmodelscorruption}. Thus, from an attacker's perspective, fine-tuning an ESD-erased model to reintroduce the erased concept is essentially equivalent to fine-tuning a standard SD model.

In contrast, immunization methods (\eg IMMA) cause fine-tuning to continually produce pure noise,  preventing the re-emergence of the concept. However, as discussed earlier in Section~\ref{sec:objects} and further in Section~\ref{sec:nsfw}, IMMA significantly degrades model performance on unrelated, safe concepts.

GIFT prevents Van Gogh-style generation entirely up to approximately step 600. Beyond this point, GIFT produces results that lie in a sweet spot between those of erasure-based (ESD) and immunization-based (IMMA) methods. Notably, GIFT allows limited re-learning from the data, which is beneficial when fine-tuning on benign inputs. The model appears to map prompts to training images, but it does not recover the generalizable ability to generate in the artist’s style. This is evident from the outputs in Fig.~\ref{fig:van-gogh-comparison}, where generated images closely resemble training examples but fail to match the prompt, resulting in lower CLIP scores. The slight increase in CLIP reflects that a Van Gogh-like image is produced, but it does not align with the intended subject (\eg a cat). 

\begin{figure}[t]
    \centering
    \includegraphics[width=1\linewidth]{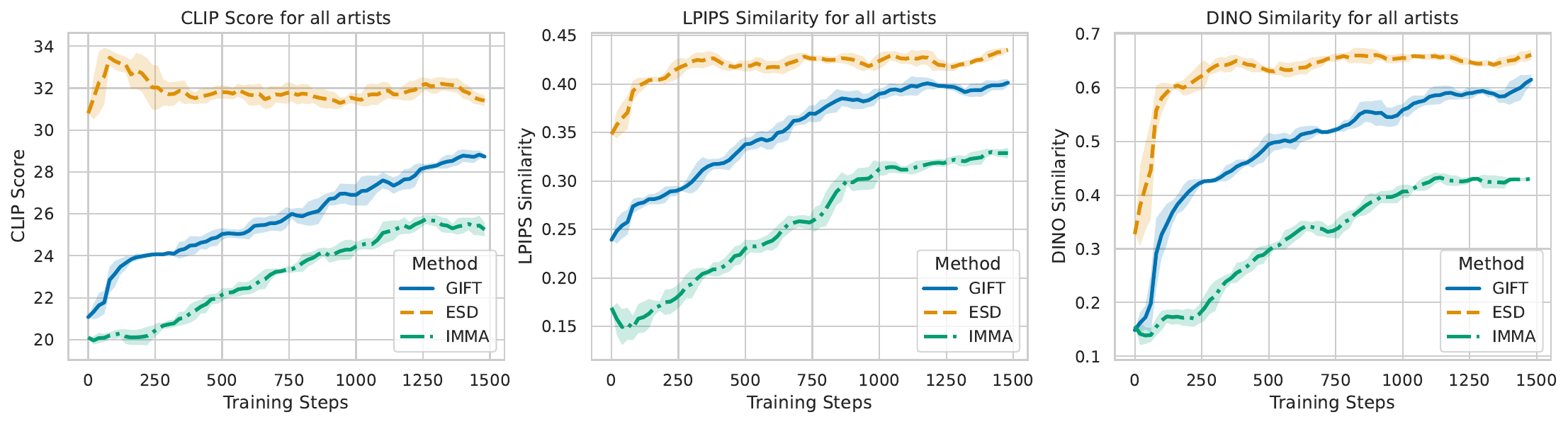}
    \caption{{\bf Quantitative Results for All Artists.} Comparison of CLIP Score, LPIPS Similarity, and DINO Similarity over fine-tuning steps for all evaluated artists. GIFT maintains lower LPIPS and DINO similarity than ESD, indicating reduced memorization of training data. The CLIP score remains substantially lower for GIFT, demonstrating its effectiveness in preventing prompt-consistent generation of the protected artistic styles.}
    \label{fig:avg-artist-comparison}
\end{figure}

This overall trend holds across all evaluated artists, as illustrated in Fig.~\ref{fig:avg-artist-comparison}. GIFT consistently yields generations with lower LPIPS and DINO similarity compared to ESD, indicating reduced memorization and less precise replication of the training data.  GIFT  exhibits slightly higher similarity than IMMA, due to its capacity to overfit on individual samples without fully re-acquiring the erased concept. Despite  this, GIFT fails to produce prompt-aligned generations throughout the training process, as evidenced by the much lower CLIP scores. This confirms that while GIFT permits limited data memorization, it successfully impedes the model from regaining the protected artistic style.


\subsection{NSFW Content}
\label{sec:nsfw}

\begin{figure}[t]
    \centering
    \includegraphics[width=1\linewidth]{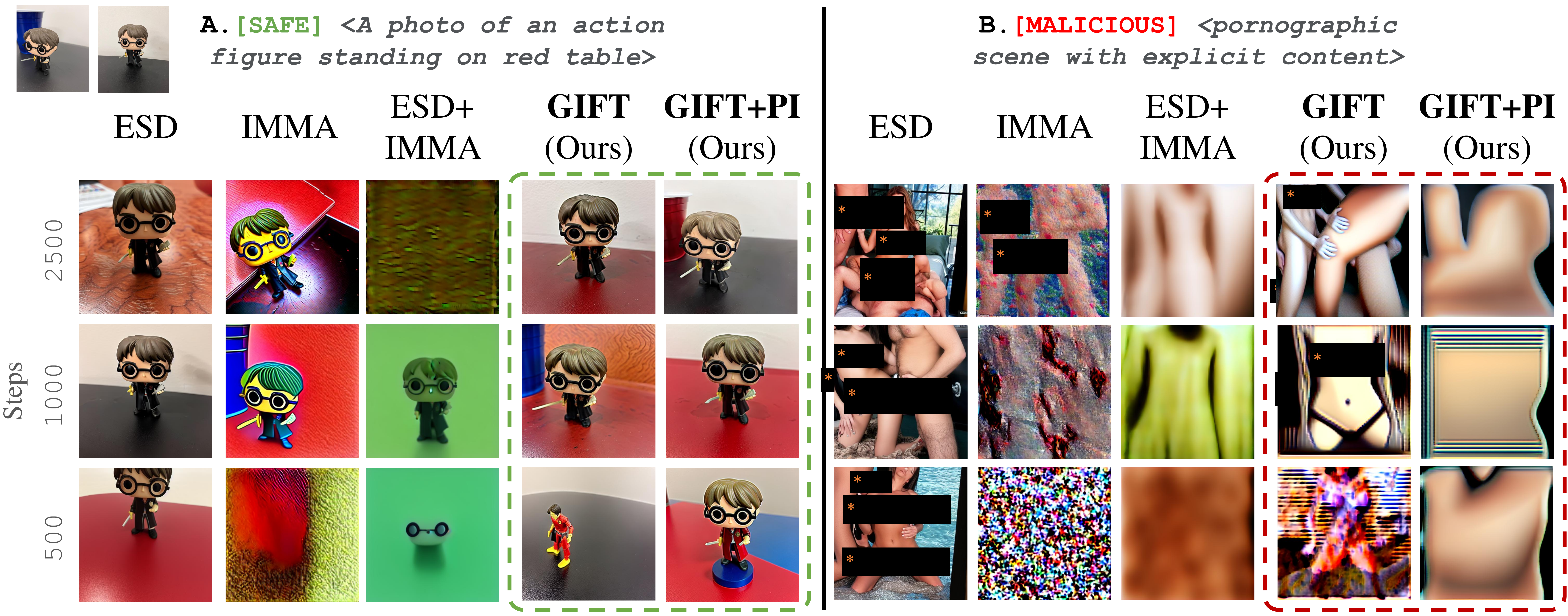}
    \caption{{\bf GIFT Blocks Malicious Fine-Tuning While Preserving Safe Adaptation.} We fine-tune each NSFW-immune model on a safe concept \textbf{(A)} and a malicious one \textbf{(B)}, evaluating at steps 500, 1000, and 2500. ESD permits safe learning but fails to block unsafe content; IMMA blocks unsafe content but harms safe fidelity; IMMA+ESD fails at both. GIFT retains safe learning while resisting malicious adaptation. GIFT+PI further improves both.}

    \label{fig:safe-malicious}
\end{figure}



\textbf{Attack Results.} During the malicious fine-tuning attack, we observe that ESD quickly allows the model to recover explicit content. IMMA prevents re-learning but does so by significantly degrading the model's learning ability across all concepts, not just NSFW. In contrast, GIFT consistently suppresses such malicious adaptation, yielding noisy or failed generations when prompted with NSFW content. It does so while preserving the ability to learn safe concepts as shown in Fig.~\ref{fig:safe-malicious}. To further enhance performance on safe concepts, we apply a post-immunization (PI) fine-tuning step to the GIFT-immunized model. This involves training on a generic, benign prompt (\eg \texttt{<A photo of a barn and mountains>}) for 1000 steps. Interestingly, this additional step not only improves the model’s ability to retain safe generation quality but also strengthens its resistance to malicious NSFW re-adaptation. We leave a deeper investigation of this effect to future work. These results demonstrate GIFT’s ability to impose robust and persistent resistance to harmful concept injection without compromising general generation quality.

\section{Additional Analysis}
\begin{figure}[t]
    \centering
    \includegraphics[width=0.95\linewidth]{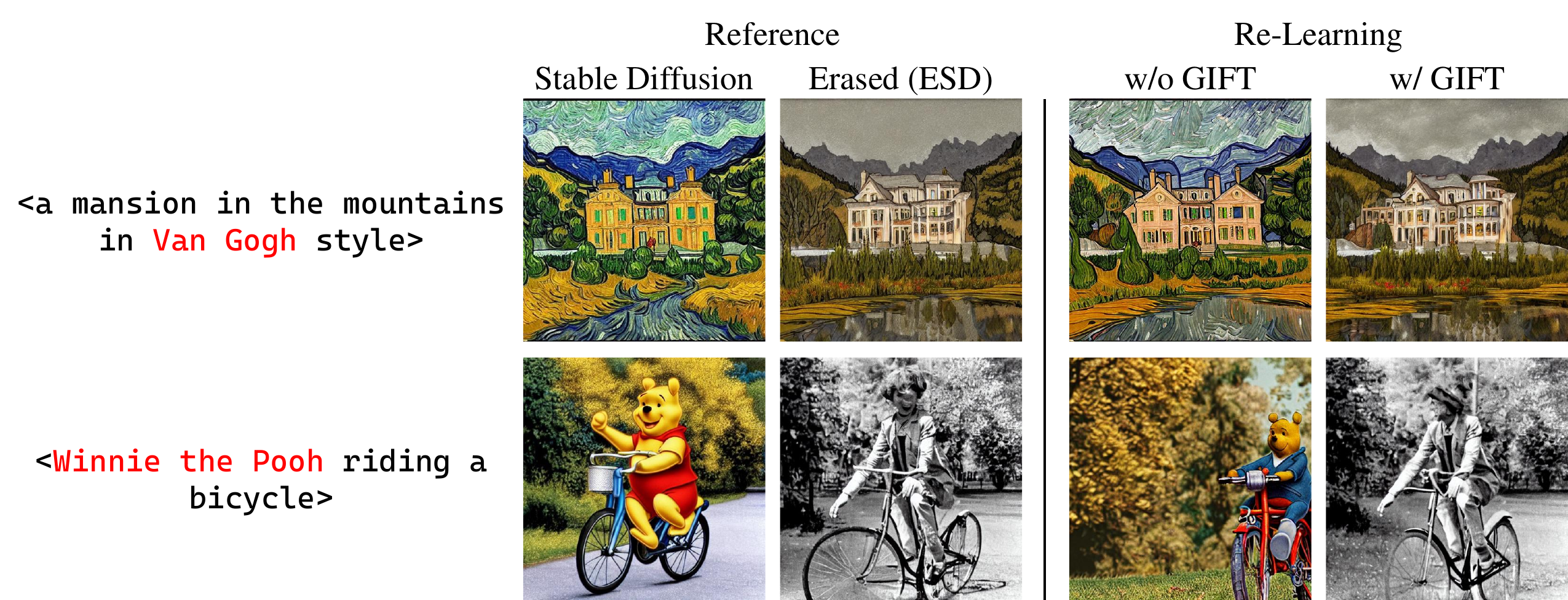}
    \caption{{\bf GIFT Immunization with LoRA.} GIFT can prevent model adaption using LoRA. }
    \label{fig:lora}
\end{figure}

\textbf{Attacking with Different Adaptation Method.}
Prior methods like IMMA perform a separate immunization process for each attack technique: immunize with DreamBooth to protect against DreamBooth (resp.~LoRA) fine-tuning. GIFT, on the other hand, does not depend on the attack algorithm during immunization. We show in Fig.~\ref{fig:lora} how our model performs with the same immunization technique used in prior sections against a different attack method, namely LoRA. We fine-tune a LoRA adapter once using an erased Stable Diffusion v1.5 (ESD) and once using our own immunized model. Fig.~\ref{fig:lora} shows that an adapter trained using an un-immunized model can easily re-acquire erased knowledge, while using GIFT it cannot.

\section{Limitations and Negative Impacts}
\label{sec:limitations_and_ethical_considerations}
While GIFT effectively immunizes text-to-image diffusion models against malicious fine-tuning, several limitations remain. First, our approach assumes access to clearly defined and representative unsafe concept datasets. In real world, such representative datasets may be hard to curate. Second, our immunization loss may still impact generation quality for safe concepts, especially when visual features overlap between safe and unsafe categories. Additionally, we focus on single concept immunization in this work. Multi-concept immunization will be explored in future work.  

From an ethical standpoint, our method is designed to reduce the risk of generating harmful, unsafe, or copyrighted content. However, it does not guarantee full immunity and could potentially be circumvented by future, more sophisticated adaptation techniques. As with any content moderation tool, misuse or overreach (\eg censoring legitimate creative expression) remains a concern. We encourage the community to treat GIFT as a step toward safer generative models, not a definitive solution, and to accompany its use with broader societal oversight.

\section{Conclusion}

This paper introduces GIFT, a gradient-aware immunization framework for diffusion models, which addresses the critical vulnerability of diffusion models to malicious fine-tuning. While previously developed safety mechanisms either degrade overall model performance (\eg IMMA) or can be easily circumvented (\eg ESD), GIFT strikes a balance between immunization effectiveness and preservation of generative capabilities on safe concepts. We formulate immunization as a bi-level optimization problem: the lower-level task focuses on preserving performance on safe concepts, while the upper-level task prevents adaptation to harmful content through a combination of loss maximization and representation noising. Extensive experiments across diverse concepts show that GIFT resists harmful re-learning, maintains generation quality, and remains fine-tunable on safe data. This makes GIFT a practical tool for safer model deployment. Future work will explore multi-concept immunization, efficient scaling, and broader application to other generative architectures. While GIFT is a key step toward model safety, it should be complemented by policy and ethical oversight for responsible AI deployment.
\clearpage
\bibliographystyle{plainnat}
\bibliography{references}

\begin{thebibliography}{39}
\providecommand{\natexlab}[1]{#1}
\providecommand{\url}[1]{\texttt{#1}}
\expandafter\ifx\csname urlstyle\endcsname\relax
  \providecommand{\doi}[1]{doi: #1}\else
  \providecommand{\doi}{doi: \begingroup \urlstyle{rm}\Url}\fi

\bibitem[Alberti et~al.(2025)Alberti, Hasanaliyev, Shah, and Ermon]{alberti2025data}
Silas Alberti, Kenan Hasanaliyev, Manav Shah, and Stefano Ermon.
\newblock Data unlearning in diffusion models.
\newblock In \emph{The Thirteenth International Conference on Learning Representations}, 2025.
\newblock URL \url{https://openreview.net/forum?id=SuHScQv5gP}.

\bibitem[Bedapudi(2019)]{bedapudi2019nudenet}
Praneet Bedapudi.
\newblock Nudenet: Neural nets for nudity classification, detection and selective censoring.
\newblock \url{https://github.com/platelminto/NudeNetClassifier}, 2019.

\bibitem[CompVis(2022)]{compvis2022stable}
CompVis.
\newblock Stable diffusion license, 2022.
\newblock URL \url{https://github.com/CompVis/stable-diffusion/blob/main/LICENSE}.
\newblock CreativeML OpenRAIL-M License.

\bibitem[Finn et~al.(2017)Finn, Abbeel, and Levine]{finn2017model}
Chelsea Finn, Pieter Abbeel, and Sergey Levine.
\newblock Model-agnostic meta-learning for fast adaptation of deep networks.
\newblock In \emph{International conference on machine learning}, pages 1126--1135. PMLR, 2017.

\bibitem[Gal et~al.(2023)Gal, Alaluf, Atzmon, Patashnik, Bermano, Chechik, and Cohen-or]{gal2023an}
Rinon Gal, Yuval Alaluf, Yuval Atzmon, Or~Patashnik, Amit~Haim Bermano, Gal Chechik, and Daniel Cohen-or.
\newblock An image is worth one word: Personalizing text-to-image generation using textual inversion.
\newblock In \emph{The Eleventh International Conference on Learning Representations}, 2023.
\newblock URL \url{https://openreview.net/forum?id=NAQvF08TcyG}.

\bibitem[Gandikota et~al.(2023)Gandikota, Materzynska, Fiotto-Kaufman, and Bau]{gandikota2023erasing}
Rohit Gandikota, Joanna Materzynska, Jaden Fiotto-Kaufman, and David Bau.
\newblock Erasing concepts from diffusion models.
\newblock In \emph{Proceedings of the IEEE/CVF International Conference on Computer Vision (ICCV)}, pages 2426--2436, October 2023.

\bibitem[Gandikota et~al.(2024)Gandikota, Orgad, Belinkov, Materzy\'nska, and Bau]{gandikota2024unified}
Rohit Gandikota, Hadas Orgad, Yonatan Belinkov, Joanna Materzy\'nska, and David Bau.
\newblock Unified concept editing in diffusion models.
\newblock In \emph{Proceedings of the IEEE/CVF Winter Conference on Applications of Computer Vision (WACV)}, pages 5111--5120, January 2024.

\bibitem[Gao et~al.(2024)Gao, Jia, Huang, Duan, Gu, Bai, Liu, and Guo]{gao2024htsattackheuristictokensearch}
Sensen Gao, Xiaojun Jia, Yihao Huang, Ranjie Duan, Jindong Gu, Yang Bai, Yang Liu, and Qing Guo.
\newblock Hts-attack: Heuristic token search for jailbreaking text-to-image models, 2024.
\newblock URL \url{https://arxiv.org/abs/2408.13896}.

\bibitem[Gong et~al.(2024)Gong, Chen, Wei, Chen, and Jiang]{rece_chao_gong}
Chao Gong, Kai Chen, Zhipeng Wei, Jingjing Chen, and Yu-Gang Jiang.
\newblock Reliable and efficient concept erasure of text-to-image diffusion models.
\newblock In Ale{\v{s}} Leonardis, Elisa Ricci, Stefan Roth, Olga Russakovsky, Torsten Sattler, and G{\"u}l Varol, editors, \emph{Computer Vision -- ECCV 2024}, pages 73--88, Cham, 2024. Springer Nature Switzerland.
\newblock ISBN 978-3-031-73668-1.

\bibitem[Hessel et~al.(2021)Hessel, Holtzman, Forbes, Bras, and Choi]{Hessel2021CLIPScoreAR}
Jack Hessel, Ari Holtzman, Maxwell Forbes, Ronan~Le Bras, and Yejin Choi.
\newblock Clipscore: A reference-free evaluation metric for image captioning.
\newblock \emph{ArXiv}, abs/2104.08718, 2021.
\newblock URL \url{https://api.semanticscholar.org/CorpusID:233296711}.

\bibitem[Hu et~al.(2022)Hu, yelong shen, Wallis, Allen-Zhu, Li, Wang, Wang, and Chen]{hu2022lora}
Edward~J Hu, yelong shen, Phillip Wallis, Zeyuan Allen-Zhu, Yuanzhi Li, Shean Wang, Lu~Wang, and Weizhu Chen.
\newblock Lo{RA}: Low-rank adaptation of large language models.
\newblock In \emph{International Conference on Learning Representations}, 2022.
\newblock URL \url{https://openreview.net/forum?id=nZeVKeeFYf9}.

\bibitem[Imagen-Team-Google et~al.(2024)Imagen-Team-Google, :, Baldridge, Bauer, Bhutani, Brichtova, Bunner, Castrejon, Chan, Chen, Dieleman, Du, Eaton-Rosen, Fei, de~Freitas, Gao, Gladchenko, Colmenarejo, Guo, Haig, Hawkins, Hu, Huang, Igwe, Kaplanis, Khodadadeh, Kim, Konyushkova, Langner, Lau, Lawton, Luo, Mokrá, Nandwani, Onoe, van~den Oord, Parekh, Pont-Tuset, Qi, Qian, Ramachandran, Rane, Rashwan, Razavi, Riachi, Srinivasan, Srinivasan, Strudel, Uria, Wang, Wang, Waters, Wolff, Wright, Xiao, Xiong, Xu, van Zee, Zhang, Zhang, Zhou, Zolna, Aboubakar, Akbulut, Akerlund, Albuquerque, Anderson, Andreetto, Aroyo, Bariach, Barker, Ben, Berman, Biles, Blok, Botadra, Brennan, Brown, Buckley, Bunel, Bursztein, Butterfield, Caine, Carpenter, Casagrande, Chang, Chang, Chaudhuri, Chen, Choi, Churbanau, Clement, Cohen, Cole, Dektiarev, Du, Dutta, Eccles, Elue, Feden, Fruchter, Garcia, Garg, Ge, Ghazy, Gipson, Goodman, Górny, Gowal, Gupta, Halpern, Han, Hao, Hayes, Heek, Hertz, Hirst, Hoogeboom, Hou, Howard, Ibrahim,
  Ike-Njoku, Iljazi, Ionescu, Isaac, Jana, Jennings, Jenson, Jia, Jones, Ju, Kajic, Kaplanis, Ayan, Kelly, Kothawade, Kouridi, Ktena, Kumakaw, Kurniawan, Lagun, Lavitas, Lee, Li, Liang, Li-Calis, Liu, Alberca, Lorrain, Lu, Lum, Ma, Malik, Mellor, Mensink, Mosseri, Murray, Nematzadeh, Nicholas, Nørly, Oliveira, Ortiz-Jimenez, Paganini, Paine, Paiss, Parrish, Peckham, Peswani, Petrovski, Pfaff, Pirozhenko, Poplin, Prabhu, Qi, Rahtz, Rashtchian, Rastogi, Raul, Razavi, Rebuffi, Ricco, Riedel, Robinson, Rohatgi, Rosgen, Rumbley, Ryu, Salgado, Salimans, Singla, Schroff, Schumann, Shah, Shaw, Shaw, Shillingford, Shivakumar, Shtatnov, Singer, Sluzhaev, Sokolov, Sottiaux, Stimberg, Stone, Stutz, Su, Tabellion, Tang, Tao, Thomas, Thornton, Toor, Udrescu, Upadhyay, Vasconcelos, Vasiloff, Voynov, Walker, Wang, Wang, Wang, Wang, Wang, Wang, Ágoston Weisz, Wiles, Wu, Xu, Xue, Yang, Yu, Yurtoglu, Zand, Zhang, Zhang, Zhao, Zhaxybay, Zhou, Zhu, Zhu, Bloxwich, Bordbar, Cobo, Collins, Dai, Doshi, Dragan, Eck, Hassabis, Hsiao,
  Hume, Kavukcuoglu, King, Krawczyk, Li, Meier-Hellstern, Orban, Pinsky, Subramanya, Vinyals, Yu, and Zwols]{imagenteamgoogle2024imagen3}
Imagen-Team-Google, :, Jason Baldridge, Jakob Bauer, Mukul Bhutani, Nicole Brichtova, Andrew Bunner, Lluis Castrejon, Kelvin Chan, Yichang Chen, Sander Dieleman, Yuqing Du, Zach Eaton-Rosen, Hongliang Fei, Nando de~Freitas, Yilin Gao, Evgeny Gladchenko, Sergio~Gómez Colmenarejo, Mandy Guo, Alex Haig, Will Hawkins, Hexiang Hu, Huilian Huang, Tobenna~Peter Igwe, Christos Kaplanis, Siavash Khodadadeh, Yelin Kim, Ksenia Konyushkova, Karol Langner, Eric Lau, Rory Lawton, Shixin Luo, Soňa Mokrá, Henna Nandwani, Yasumasa Onoe, Aäron van~den Oord, Zarana Parekh, Jordi Pont-Tuset, Hang Qi, Rui Qian, Deepak Ramachandran, Poorva Rane, Abdullah Rashwan, Ali Razavi, Robert Riachi, Hansa Srinivasan, Srivatsan Srinivasan, Robin Strudel, Benigno Uria, Oliver Wang, Su~Wang, Austin Waters, Chris Wolff, Auriel Wright, Zhisheng Xiao, Hao Xiong, Keyang Xu, Marc van Zee, Junlin Zhang, Katie Zhang, Wenlei Zhou, Konrad Zolna, Ola Aboubakar, Canfer Akbulut, Oscar Akerlund, Isabela Albuquerque, Nina Anderson, Marco Andreetto, Lora
  Aroyo, Ben Bariach, David Barker, Sherry Ben, Dana Berman, Courtney Biles, Irina Blok, Pankil Botadra, Jenny Brennan, Karla Brown, John Buckley, Rudy Bunel, Elie Bursztein, Christina Butterfield, Ben Caine, Viral Carpenter, Norman Casagrande, Ming-Wei Chang, Solomon Chang, Shamik Chaudhuri, Tony Chen, John Choi, Dmitry Churbanau, Nathan Clement, Matan Cohen, Forrester Cole, Mikhail Dektiarev, Vincent Du, Praneet Dutta, Tom Eccles, Ndidi Elue, Ashley Feden, Shlomi Fruchter, Frankie Garcia, Roopal Garg, Weina Ge, Ahmed Ghazy, Bryant Gipson, Andrew Goodman, Dawid Górny, Sven Gowal, Khyatti Gupta, Yoni Halpern, Yena Han, Susan Hao, Jamie Hayes, Jonathan Heek, Amir Hertz, Ed~Hirst, Emiel Hoogeboom, Tingbo Hou, Heidi Howard, Mohamed Ibrahim, Dirichi Ike-Njoku, Joana Iljazi, Vlad Ionescu, William Isaac, Reena Jana, Gemma Jennings, Donovon Jenson, Xuhui Jia, Kerry Jones, Xiaoen Ju, Ivana Kajic, Christos Kaplanis, Burcu~Karagol Ayan, Jacob Kelly, Suraj Kothawade, Christina Kouridi, Ira Ktena, Jolanda Kumakaw, Dana
  Kurniawan, Dmitry Lagun, Lily Lavitas, Jason Lee, Tao Li, Marco Liang, Maggie Li-Calis, Yuchi Liu, Javier~Lopez Alberca, Matthieu~Kim Lorrain, Peggy Lu, Kristian Lum, Yukun Ma, Chase Malik, John Mellor, Thomas Mensink, Inbar Mosseri, Tom Murray, Aida Nematzadeh, Paul Nicholas, Signe Nørly, João~Gabriel Oliveira, Guillermo Ortiz-Jimenez, Michela Paganini, Tom~Le Paine, Roni Paiss, Alicia Parrish, Anne Peckham, Vikas Peswani, Igor Petrovski, Tobias Pfaff, Alex Pirozhenko, Ryan Poplin, Utsav Prabhu, Yuan Qi, Matthew Rahtz, Cyrus Rashtchian, Charvi Rastogi, Amit Raul, Ali Razavi, Sylvestre-Alvise Rebuffi, Susanna Ricco, Felix Riedel, Dirk Robinson, Pankaj Rohatgi, Bill Rosgen, Sarah Rumbley, Moonkyung Ryu, Anthony Salgado, Tim Salimans, Sahil Singla, Florian Schroff, Candice Schumann, Tanmay Shah, Eleni Shaw, Gregory Shaw, Brendan Shillingford, Kaushik Shivakumar, Dennis Shtatnov, Zach Singer, Evgeny Sluzhaev, Valerii Sokolov, Thibault Sottiaux, Florian Stimberg, Brad Stone, David Stutz, Yu-Chuan Su, Eric
  Tabellion, Shuai Tang, David Tao, Kurt Thomas, Gregory Thornton, Andeep Toor, Cristian Udrescu, Aayush Upadhyay, Cristina Vasconcelos, Alex Vasiloff, Andrey Voynov, Amanda Walker, Luyu Wang, Miaosen Wang, Simon Wang, Stanley Wang, Qifei Wang, Yuxiao Wang, Ágoston Weisz, Olivia Wiles, Chenxia Wu, Xingyu~Federico Xu, Andrew Xue, Jianbo Yang, Luo Yu, Mete Yurtoglu, Ali Zand, Han Zhang, Jiageng Zhang, Catherine Zhao, Adilet Zhaxybay, Miao Zhou, Shengqi Zhu, Zhenkai Zhu, Dawn Bloxwich, Mahyar Bordbar, Luis~C. Cobo, Eli Collins, Shengyang Dai, Tulsee Doshi, Anca Dragan, Douglas Eck, Demis Hassabis, Sissie Hsiao, Tom Hume, Koray Kavukcuoglu, Helen King, Jack Krawczyk, Yeqing Li, Kathy Meier-Hellstern, Andras Orban, Yury Pinsky, Amar Subramanya, Oriol Vinyals, Ting Yu, and Yori Zwols.
\newblock Imagen 3, 2024.
\newblock URL \url{https://arxiv.org/abs/2408.07009}.

\bibitem[Kumari et~al.(2023{\natexlab{a}})Kumari, Zhang, Wang, Shechtman, Zhang, and Zhu]{kumari2023conceptablation}
Nupur Kumari, Bingliang Zhang, Sheng-Yu Wang, Eli Shechtman, Richard Zhang, and Jun-Yan Zhu.
\newblock Ablating concepts in text-to-image diffusion models.
\newblock In \emph{Proceedings of the IEEE/CVF International Conference on Computer Vision (ICCV)}, pages 22691--22702, October 2023{\natexlab{a}}.

\bibitem[Kumari et~al.(2023{\natexlab{b}})Kumari, Zhang, Zhang, Shechtman, and Zhu]{Kumari_2023_CVPR}
Nupur Kumari, Bingliang Zhang, Richard Zhang, Eli Shechtman, and Jun-Yan Zhu.
\newblock Multi-concept customization of text-to-image diffusion.
\newblock In \emph{Proceedings of the IEEE/CVF Conference on Computer Vision and Pattern Recognition (CVPR)}, pages 1931--1941, June 2023{\natexlab{b}}.

\bibitem[Liu et~al.(2024{\natexlab{a}})Liu, Wang, Cao, Jia, and Huang]{Liu_2024_CVPR}
Bingyan Liu, Chengyu Wang, Tingfeng Cao, Kui Jia, and Jun Huang.
\newblock Towards understanding cross and self-attention in stable diffusion for text-guided image editing.
\newblock In \emph{Proceedings of the IEEE/CVF Conference on Computer Vision and Pattern Recognition (CVPR)}, pages 7817--7826, June 2024{\natexlab{a}}.

\bibitem[Liu et~al.(2024{\natexlab{b}})Liu, Cui, Li, Li, Huang, Xia, Zhang, Zou, and He]{liu2024jailbreak}
Xuannan Liu, Xing Cui, Peipei Li, Zekun Li, Huaibo Huang, Shuhan Xia, Miaoxuan Zhang, Yueying Zou, and Ran He.
\newblock Jailbreak attacks and defenses against multimodal generative models: A survey.
\newblock \emph{arXiv preprint arXiv:2411.09259}, 2024{\natexlab{b}}.

\bibitem[Patel and Qiu(2025)]{patel2025learningunlearnretainingcombating}
Gaurav Patel and Qiang Qiu.
\newblock Learning to unlearn while retaining: Combating gradient conflicts in machine unlearning, 2025.
\newblock URL \url{https://arxiv.org/abs/2503.06339}.

\bibitem[Pham et~al.(2024)Pham, Marshall, Cohen, Mittal, and Hegde]{pham2024circumventing}
Minh Pham, Kelly~O. Marshall, Niv Cohen, Govind Mittal, and Chinmay Hegde.
\newblock Circumventing concept erasure methods for text-to-image generative models.
\newblock In \emph{The Twelfth International Conference on Learning Representations}, 2024.
\newblock URL \url{https://openreview.net/forum?id=ag3o2T51Ht}.

\bibitem[Podell et~al.(2023)Podell, English, Lacey, Blattmann, Dockhorn, Müller, Penna, and Rombach]{podell2023sdxlimprovinglatentdiffusion}
Dustin Podell, Zion English, Kyle Lacey, Andreas Blattmann, Tim Dockhorn, Jonas Müller, Joe Penna, and Robin Rombach.
\newblock Sdxl: Improving latent diffusion models for high-resolution image synthesis, 2023.
\newblock URL \url{https://arxiv.org/abs/2307.01952}.

\bibitem[Ramesh et~al.(2021)Ramesh, Pavlov, Goh, Gray, Voss, Radford, Chen, and Sutskever]{dalle_ramesh_2021}
Aditya Ramesh, Mikhail Pavlov, Gabriel Goh, Scott Gray, Chelsea Voss, Alec Radford, Mark Chen, and Ilya Sutskever.
\newblock Zero-shot text-to-image generation.
\newblock In Marina Meila and Tong Zhang, editors, \emph{Proceedings of the 38th International Conference on Machine Learning}, volume 139 of \emph{Proceedings of Machine Learning Research}, pages 8821--8831. PMLR, 18--24 Jul 2021.
\newblock URL \url{https://proceedings.mlr.press/v139/ramesh21a.html}.

\bibitem[Rando et~al.(2022)Rando, Paleka, Lindner, Heim, and Tram{\`e}r]{rando2022red}
Javier Rando, Daniel Paleka, David Lindner, Lennart Heim, and Florian Tram{\`e}r.
\newblock Red-teaming the stable diffusion safety filter.
\newblock \emph{arXiv preprint arXiv:2210.04610}, 2022.

\bibitem[Rombach et~al.(2022)Rombach, Blattmann, Lorenz, Esser, and Ommer]{Rombach_2022_CVPR}
Robin Rombach, Andreas Blattmann, Dominik Lorenz, Patrick Esser, and Bj\"orn Ommer.
\newblock High-resolution image synthesis with latent diffusion models.
\newblock In \emph{Proceedings of the IEEE/CVF Conference on Computer Vision and Pattern Recognition (CVPR)}, pages 10684--10695, June 2022.

\bibitem[Rosati et~al.(2024)Rosati, Wehner, Williams, Bartoszcze, Atanasov, Gonzales, Majumdar, Maple, Sajjad, and Rudzicz]{rosati2024representation}
Domenic Rosati, Jan Wehner, Kai Williams, \L~ukasz Bartoszcze, David Atanasov, Robie Gonzales, Subhabrata Majumdar, Carsten Maple, Hassan Sajjad, and Frank Rudzicz.
\newblock Representation noising: A defence mechanism against harmful finetuning.
\newblock In A.~Globerson, L.~Mackey, D.~Belgrave, A.~Fan, U.~Paquet, J.~Tomczak, and C.~Zhang, editors, \emph{Advances in Neural Information Processing Systems}, volume~37, pages 12636--12676. Curran Associates, Inc., 2024.

\bibitem[Ruiz et~al.(2023)Ruiz, Li, Jampani, Pritch, Rubinstein, and Aberman]{Ruiz_2023_CVPR}
Nataniel Ruiz, Yuanzhen Li, Varun Jampani, Yael Pritch, Michael Rubinstein, and Kfir Aberman.
\newblock Dreambooth: Fine tuning text-to-image diffusion models for subject-driven generation.
\newblock In \emph{Proceedings of the IEEE/CVF Conference on Computer Vision and Pattern Recognition (CVPR)}, pages 22500--22510, June 2023.

\bibitem[Schramowski et~al.(2023)Schramowski, Brack, Deiseroth, and Kersting]{schramowski2023safe}
Patrick Schramowski, Manuel Brack, Bj{\"o}rn Deiseroth, and Kristian Kersting.
\newblock Safe latent diffusion: Mitigating inappropriate degeneration in diffusion models.
\newblock In \emph{Proceedings of the IEEE/CVF Conference on Computer Vision and Pattern Recognition}, pages 22522--22531, 2023.

\bibitem[Shan et~al.(2023)Shan, Cryan, Wenger, Zheng, Hanocka, and Zhao]{shawn_2023}
Shawn Shan, Jenna Cryan, Emily Wenger, Haitao Zheng, Rana Hanocka, and Ben~Y. Zhao.
\newblock Glaze: protecting artists from style mimicry by text-to-image models.
\newblock In \emph{Proceedings of the 32nd USENIX Conference on Security Symposium}, SEC '23, USA, 2023. USENIX Association.
\newblock ISBN 978-1-939133-37-3.

\bibitem[Tian et~al.(2024)Tian, Liang, Cheng, Liu, Wang, Sui, Chen, Chen, and Zhang]{tian-etal-2024-forget}
Bozhong Tian, Xiaozhuan Liang, Siyuan Cheng, Qingbin Liu, Mengru Wang, Dianbo Sui, Xi~Chen, Huajun Chen, and Ningyu Zhang.
\newblock To forget or not? towards practical knowledge unlearning for large language models.
\newblock In Yaser Al-Onaizan, Mohit Bansal, and Yun-Nung Chen, editors, \emph{Findings of the Association for Computational Linguistics: EMNLP 2024}, pages 1524--1537, Miami, Florida, USA, November 2024. Association for Computational Linguistics.
\newblock \doi{10.18653/v1/2024.findings-emnlp.82}.
\newblock URL \url{https://aclanthology.org/2024.findings-emnlp.82/}.

\bibitem[Wang et~al.(2024)Wang, Chen, Li, Yang, and Ji]{wang2024aeiouunifieddefenseframework}
Yiming Wang, Jiahao Chen, Qingming Li, Xing Yang, and Shouling Ji.
\newblock Aeiou: A unified defense framework against nsfw prompts in text-to-image models, 2024.
\newblock URL \url{https://arxiv.org/abs/2412.18123}.

\bibitem[Wu et~al.(2024)Wu, Zhang, Hua, Lyu, Wang, Song, and Guan]{wu2024exploringdiffusionmodelscorruption}
Xiaoyu Wu, Jiaru Zhang, Yang Hua, Bohan Lyu, Hao Wang, Tao Song, and Haibing Guan.
\newblock Exploring diffusion models' corruption stage in few-shot fine-tuning and mitigating with bayesian neural networks, 2024.
\newblock URL \url{https://arxiv.org/abs/2405.19931}.

\bibitem[Wu et~al.(2025)Wu, Zhou, Yang, Wang, Chang, Zhu, Hu, Zhou, and Yang]{wu2025unlearning}
Yongliang Wu, Shiji Zhou, Mingzhuo Yang, Lianzhe Wang, Heng Chang, Wenbo Zhu, Xinting Hu, Xiao Zhou, and Xu~Yang.
\newblock Unlearning concepts in diffusion model via concept domain correction and concept preserving gradient.
\newblock \emph{Proceedings of the AAAI Conference on Artificial Intelligence}, 39\penalty0 (8):\penalty0 8496--8504, Apr. 2025.
\newblock \doi{10.1609/aaai.v39i8.32917}.
\newblock URL \url{https://ojs.aaai.org/index.php/AAAI/article/view/32917}.

\bibitem[Xu et~al.(2023)Xu, Zhu, Zhang, Zhou, and Yu]{Xu_MU_Survey}
Heng Xu, Tianqing Zhu, Lefeng Zhang, Wanlei Zhou, and Philip~S. Yu.
\newblock Machine unlearning: A survey.
\newblock \emph{ACM Comput. Surv.}, 56\penalty0 (1), August 2023.
\newblock ISSN 0360-0300.
\newblock \doi{10.1145/3603620}.
\newblock URL \url{https://doi.org/10.1145/3603620}.

\bibitem[Yang et~al.(2024)Yang, Hui, Yuan, Gong, and Cao]{yang_sneaky_2024}
Yuchen Yang, Bo~Hui, Haolin Yuan, Neil Gong, and Yinzhi Cao.
\newblock Sneakyprompt: Jailbreaking text-to-image generative models.
\newblock In \emph{2024 IEEE Symposium on Security and Privacy (SP)}, pages 897--912, 2024.
\newblock \doi{10.1109/SP54263.2024.00123}.

\bibitem[Yoon et~al.(2025)Yoon, Yu, Patil, Yao, and Bansal]{yoon2025safree}
Jaehong Yoon, Shoubin Yu, Vaidehi Patil, Huaxiu Yao, and Mohit Bansal.
\newblock {SAFREE}: Training-free and adaptive guard for safe text-to-image and video generation.
\newblock In \emph{The Thirteenth International Conference on Learning Representations}, 2025.
\newblock URL \url{https://openreview.net/forum?id=hgTFotBRKl}.

\bibitem[Zhang et~al.(2024{\natexlab{a}})Zhang, Wang, Xu, Wang, and Shi]{Zhang_2024_CVPR}
Gong Zhang, Kai Wang, Xingqian Xu, Zhangyang Wang, and Humphrey Shi.
\newblock Forget-me-not: Learning to forget in text-to-image diffusion models.
\newblock In \emph{Proceedings of the IEEE/CVF Conference on Computer Vision and Pattern Recognition (CVPR) Workshops}, pages 1755--1764, June 2024{\natexlab{a}}.

\bibitem[Zhang et~al.(2018)Zhang, Isola, Efros, Shechtman, and Wang]{zhang2018unreasonable}
Richard Zhang, Phillip Isola, Alexei~A Efros, Eli Shechtman, and Oliver Wang.
\newblock The unreasonable effectiveness of deep features as a perceptual metric.
\newblock In \emph{Proceedings of the IEEE Conference on Computer Vision and Pattern Recognition}, pages 586--595, 2018.

\bibitem[Zhang et~al.(2024{\natexlab{b}})Zhang, Jia, Chen, Chen, Zhang, Liu, Ding, and Liu]{zhang2024generate}
Yimeng Zhang, Jinghan Jia, Xin Chen, Aochuan Chen, Yihua Zhang, Jiancheng Liu, Ke~Ding, and Sijia Liu.
\newblock To generate or not? safety-driven unlearned diffusion models are still easy to generate unsafe images... for now.
\newblock In \emph{European Conference on Computer Vision}, pages 385--403. Springer, 2024{\natexlab{b}}.

\bibitem[Zheng and Yeh(2024)]{zheng2024imma}
Amber~Yijia Zheng and Raymond~A Yeh.
\newblock Imma: Immunizing text-to-image models against malicious adaptation.
\newblock In \emph{European Conference on Computer Vision}, pages 458--475. Springer, 2024.

\bibitem[Zhou et~al.(2024)Zhou, Wang, Ye, Wu, and Chang]{Zhou2024OnTL}
Shiji Zhou, Lianzhe Wang, Jiangnan Ye, Yongliang Wu, and Heng Chang.
\newblock On the limitations and prospects of machine unlearning for generative ai.
\newblock \emph{ArXiv}, abs/2408.00376, 2024.

\bibitem[zxbsmk(2024)]{zxbsmk_nsfw_t2i}
zxbsmk.
\newblock Nsfw-t2i.
\newblock \url{https://huggingface.co/datasets/zxbsmk/NSFW-T2I}, 2024.

\end{thebibliography}



\end{document}